\documentclass[aps,pre,twocolumn,groupedaddress]{revtex4-1}

\usepackage{amsfonts,amssymb,amsmath} 
\usepackage{color} 
\usepackage{graphicx} 
\usepackage{epstopdf,mathrsfs} 
\usepackage[colorlinks,linktocpage,linkcolor=blue]{hyperref}

\begin{document}


\title{Feynman-Kac equation revisited}

\author{Xudong Wang}
\author{Yao Chen}
\author{Weihua Deng}

\affiliation{School of Mathematics and Statistics, Gansu Key Laboratory
of Applied Mathematics and Complex Systems, Lanzhou University, Lanzhou 730000,
P.R. China}



\begin{abstract}
Functionals of particles' paths have diverse applications in physics, mathematics, hydrology, economics, and other fields. Under the framework of continuous time random walk (CTRW), the governing equations for the probability density functions (PDFs) of the functionals, including the ones of the paths of stochastic processes of normal diffusion, anomalous diffusion, and even the diffusion with reaction, have been derived. Sometimes, the stochastic processes in physics and chemistry are naturally described by Langevin equations. The Langevin picture has the advantages in studying the dynamics with an external force field and analyzing the effect of noise resulting from a fluctuating environment.
We derive the governing equations of the PDFs of the functionals of paths of Langevin system with both space and time dependent force field and arbitrary multiplicative noise; and the backward version is proposed for the system with arbitrary additive noise or multiplicative Gaussian white noise together with a force field. For the newly built equations, their applications of solving the PDFs of the occupation time and area under the trajectory curve are provided, and the results are confirmed by simulations.

%
%
\end{abstract}


\maketitle

\section{Introduction}
Stochastic processes are the basic mathematical tools to describe natural phenomena. For satisfying  the demand of practical applications or understanding the microscopic mechanism, extracting statistical information of the stochastic process is one of the most important strategies. Functional is the path integration of the stochastic process, being a random variable. It has diverse applications across multi-disciplinary, ranging from probability theory \cite{Kac:1949}, mathematical finance \cite{Geman:1993}, mesocopic
physics \cite{Comtet:2005}, computer science \cite{Majumdar:2005}, and in understanding
the cooling and the heating degree days relevant to weather derivatives \cite{Majumdar:2002}. This paper focuses on deriving the governing equations of the probability density functions (PDFs) of functionals of the paths of Langevin dynamics.

The popular microscopic models of describing stochastic dynamics in the natural world include continuous time random walks (CTRWs) and Langevin equations \cite{Langevin:1908}. The Langevin picture is more convenient to apply if the effect of external field and/or noises generated from a fluctuating environment \cite{CoffeyKalmykovWaldron:2004} is considered; it builds a relation between physically transparent and mathematically tractable description for complex stochastic dynamics.
The dynamical behaviors of the system depend fundamentally on the specific form of noise. The most  common one is L\'{e}vy noise, generating L\'{e}vy process \cite{Applebaum:2009}, which is a stochastic process with stationary and independent increments and zero initial state. For L\'{e}vy noise, the solutions of Langevin equation belong to the class of Markov processes
\cite{VanKampen:1992,HorsthemkeLefever:1984,HanggiThomas:1982,Risken:1989}.
As for the governing equations of the PDFs of the displacement and/or velocity of particles described by Langevin equation, there have been well developments.
Specially, the Langevin equation with Gaussian white noise corresponds to the ordinary Fokker-Planck equation \cite{HorsthemkeLefever:1984,HanggiThomas:1982,Risken:1989} and heavy-tailed stable noise to the spatial fractional Fokker-Planck equation \cite{DenisovHorsthemkeHanggi:2008,DenisovHorsthemkeHanggi:2009,Fogedby:1994,Fogedby:1998,KolwankarGangal:1998,MetzlerBarkaiKlafter:1999,JespersenMetzlerFogedby:1999,ChechkinGoncharKlafterMetzler:2006,Adelman:1976}. Besides that, the temporal fractional Fokker-Planck equation is obtained by the time-changed Langevin equations with inverse $\alpha$-stable subordinator \cite{Magdziarz:2007}.

There are also some progresses in deriving the governing equations of the PDFs of functional: $A=\int_0^tU[x(t')]dt'$, where $x(t)$ is a path of stochastic process and $U(x)$ is some prescribed function.
Influenced by Feynman's thesis about Schr\"{o}dinger's equation, Kac derives the classical Feynman-Kac equation in 1949 for normal diffusion \cite{Kac:1949}. In recent years, Majumdar discusses the applications of Brownian functionals in \cite{Majumdar:2005} by the path integral method. Then more and more Feynman-Kac equations for non-Brownian functionals are established within the framework of CTRW models \cite{TurgemanCarmiBarkai:2009,CarmiBarkai:2011,CarmiTurgemanBarkai:2010,WuDengBarkai:2016,HouDeng:2018,WangDeng:2018,XuDeng:2017}, in particular the ones in \cite{HouDeng:2018} are for the functionals of reaction diffusion process. In some cases, by using the method of subordination \cite{Fogedby:1994}, a one-to-one  correspondence of the Langevin picture and CTRW model can be achieved, but there are still a lot of cases that Langevin picture is a more natural choice or that can not be conveniently characterized by CTRW model, e.g., the Langevin equation with multiplicative noise, being effectively used to describe the motion of amoebae \cite{Selmeczi:2008}.
Not great progresses have been made for obtaining the Feynman-Kac equations governing the PDF of the functionals of the paths of Langevin dynamics. Using the It\^{o} formula, Cairoli and Baule \cite{CairoliBaule:2015,CairoliBaule:2017} provide the derivation of the forward Feynman-Kac equation from Langevin system with Gaussian white noise and arbitrary waiting time distribution. Along this direction, by adopting some different ideas, this paper pushes forward the research of deriving the Feynman-Kac equations for more general Langevin pictures, 
for example, the dynamical system with a fluctuating environment described by the overdamped Langevin equation:
\begin{equation}\label{LEmodel}
  \dot{x}(t)=f(x(t),t)+g(x(t),t)\xi(t),
\end{equation}
where $x(t)$ is the particle coordinate, $f(x,t)$ is the force field, $\xi(t)$ is the noise resulting from a fluctuating environment, and $g(x,t)$ is the multiplicative noise term.

This paper extends the ideas in \cite{DenisovHorsthemkeHanggi:2009}, which focus on the derivation of the generalized Fokker-Planck equation, to derive the generalized Feynman-Kac equation for overdamped Langevin equation driven by an arbitrary L\'{e}vy noise together with a multiplicative noise term, then investigates applications for specific functionals of interest. To our knowledge, all the existing backward Feynman-Kac equations are obtained from CTRW models, not Langevin system, even with Gaussian white noise together with a force field.  Here we derive the backward Feynman-Kac equation from the Langevin system with multiplicative Gaussian white noise or additive arbitrary L\'{e}vy noise. 
The paper is organized as follows. In Section \ref{Sec2}, we derive the forward and backward Feynman-Kac equations associated with the overdamped Langevin equation \eqref{LEmodel}. In Section \ref{Sec3}, we use the derived equations to study two examples: the occupation time and fraction of a particle moving in a box with reflection boundary conditions, and the area under the curve of particle trajectory; and some numerical simulations are performed to verify the correctnesses of the theoretical results. Finally, the summaries are made in Section \ref{Sec4}.

\section{Derivation of the equations}\label{Sec2}
\subsection{Forward equation}

Here we use the L\'{e}vy noise $\xi(t)$, which is the formal time derivative of its corresponding L\'{e}vy process $\eta(t)$. That is to say,  the increment $\delta\eta(t)=\eta(t+\tau)-\eta(t)$ of $\eta(t)$ could be defined as the time integral of $\xi(t)$, $\delta\eta(t)=\int_t^{t+\tau}\xi(t')dt'$. Similarly, the increment $\delta x(t)=x(t+\tau)-x(t)$ of the particle trajectory undergoing the Langevin system \eqref{LEmodel} during a time interval $\tau\,(\tau\rightarrow0)$ satisfies
\begin{equation}\label{LEIto}
  \delta x(t)=f(x(t),t)\tau + g(x(t),t)\delta\eta(t),
\end{equation}
which defines the meaning of equation \eqref{LEmodel} in the It\^{o} interpretation \cite{Ito:1950,Risken:1989}.
The particle location $x(t)$ only depends on the previous increments of $\eta(t)$ and thus it is independent on the increment $\delta\eta(t)$ since the increments of L\'{e}vy process is independent on non-overlapping intervals. Because of the stationary increment of the L\'{e}vy process, we know that $\delta\eta(t)$ has the same distribution as $\eta(\tau)$ with characteristic function denoted by \cite{Applebaum:2009}:
\begin{equation}\label{NoiseChar}
  \langle e^{-ik\eta(\tau)}\rangle=e^{\tau\phi_0(k)},
\end{equation}
where the Fourier symbol $\phi_0(k)$ characterizes the jump structure of the L\'{e}vy noise $\xi(t)$. In the subsequent part, for a specific L\'{e}vy noise, it has the specific form that $\phi_0(k)=-k^2$ for Gaussian white noise and $\phi_0(k)=-|k|^\beta$ for non-Gaussian $\beta$-stable L\'{e}vy noise.

Define the functional $A=\int_0^t U[x(t')]dt'$ and $G(x,A,t)$ as the joint PDF of position $x$ and functional $A$ at time $t$.
In order to obtain the joint PDF $G(x,A,t)$, we define its Fourier transform $x\rightarrow k,~A\rightarrow p$ as:
\begin{equation*}
  G(k,p,t)=\int_{-\infty}^{\infty}\int_{-\infty}^{\infty} e^{-ikx-ipA}G(x,A,t)dxdA,
\end{equation*}
and write it in the usual way
\begin{equation}\label{Gkpt}
  G(k,p,t)=\langle e^{-ikx(t)}e^{-ipA(t)}\rangle.
\end{equation}
Through this work, we use the convention that the variables in parentheses to say what space we are working.
Being similar to the increment $\delta x(t)$ in \eqref{LEIto}, one has the increment $\delta A(t)=A(t+\tau)-A(t)=U(x(t))\tau$ during the time interval $\tau\,(\tau\rightarrow0)$.
Then we consider the increment of $G(x,A,t)$ in Fourier space, $\delta G(k,p,t):=G(k,p,t+\tau)-G(k,p,t)$, which can be written as
\begin{equation}\label{dGkpt}
  \delta G(k,p,t)=\langle e^{-ikx(t+\tau)-ipA(t+\tau)}\rangle-\langle e^{-ikx(t)-ipA(t)}\rangle.
\end{equation}
Substituting the increment $\delta x(t),\,\delta A(t)$ into \eqref{dGkpt} and taking $\tau\rightarrow0$, we obtain
\begin{equation}\label{dGkpt2}
\begin{split}
    \delta G(k,p,t)=&~ \langle e^{-ikx(t)-ipA(t)}(e^{-ikg(x(t),t)\delta\eta(t)}-1) \rangle \\
    &-ik\tau\langle e^{-ikx(t)-ipA(t)}f(x(t),t) \rangle \\
    &-ip\tau \langle e^{-ikx(t)-ipA(t)}U(x(t)) \rangle.
\end{split}
\end{equation}
Note that the angle bracket in the first term in \eqref{dGkpt2} denotes the average with the joint PDF $G(x,A,t)$ and the PDF of the noise increment $\delta\eta(t)$ since $\delta\eta(t)$ is independent of particle trajectory $x(t)$. The characteristic function of the noise increment $\delta\eta(t)$ in \eqref{NoiseChar} gives
\begin{equation}\label{Forward1}
  \lim_{\tau\rightarrow0}\frac{1}{\tau}\langle(e^{-ikg(x(t),t)\delta\eta(t)}-1)\rangle=\phi_0(kg(x(t),t)).
\end{equation}
The second and third terms in \eqref{dGkpt2} are just the Fourier transform of a compound function on $G(x,A,t)$, i.e.,
\begin{equation}\label{Forward2}
\begin{split}
 & ik\langle e^{-ikx(t)-ipA(t)}f(x(t),t) \rangle
  \\
 & =\mathcal{F}_x \mathcal{F}_A\left\{\frac{\partial}{\partial x}f(x,t)G(x,A,t)\right\},
  \end{split}
\end{equation}
and
\begin{equation}\label{Forward3}
  ip\langle e^{-ikx(t)-ipA(t)}U(x(t)) \rangle= ip\mathcal{F}_x\mathcal{F}_A\left\{U(x)G(x,A,t)\right\}.
\end{equation}
Basing on \eqref{Forward1}, \eqref{Forward2} and \eqref{Forward3}, dividing \eqref{dGkpt2} by $\tau$ and taking the limit $\tau\rightarrow0$, we obtain the forward Feynman-Kac equation in Fourier space:
\begin{equation}\label{FFKE_k}
\begin{split}
    \frac{\partial G(k,p,t)}{\partial t}&= \mathcal{F}_x\{\phi_0(kg(x,t))G(x,p,t)\} \\
     -\mathcal{F}_x\Big\{\frac{\partial}{\partial x}&f(x,t)G(x,p,t)+ipU(x)G(x,p,t)\Big\}.
\end{split}
\end{equation}
Once the form of $\phi_0(kg(x,t))$ is given for a specific noise, the forward Feynman-Kac equation in $x$ space is obtained.

If the deterministic time variable in Langevin equation \eqref{LEmodel} is replaced by a positive non-decreasing  one-dimensional L\'{e}vy process, called subordinator \cite{Applebaum:2009}, then the subordinated stochastic process could be described by the following coupled Langevin equation
\begin{equation}\label{LEmodelB}
\begin{split}
      &\dot{x}(s)=f(x(s),T(s))+g(x(s),T(s))\xi(s),  \\
      &\dot{T}(s)=\theta(s).
\end{split}
\end{equation}
Here we adopt the fully skewed $\alpha$-stable L\'{e}vy noise $\theta(s)$ with $0<\alpha<1$, which is independent of the arbitrary L\'{e}vy noise $\xi(s)$. Then the combined process is defined as $y(t)=x(S(t))$ with the inverse $\alpha$-stable subordinator $S(t)$, which is the first-passage time of the $\alpha$-stable subordinator $\{ T(s),s\geq0\}$ and defined \cite{PiryatinskaSaichevWoyczynski:2005,MagdziarzWeron:2006} as
$S(t)=\inf_{s>0}\{s:T(s)>t\}$.
Note that the time-dependent force $f$ and multiplicative noise term $g$ should depend on the physical time $T(s)$, rather than the operation time $s$, due to a physical interpretation \cite{MagdziarzWeronKlafter:2008,HeinsaluPatriarcaGoychukHanggi:2007}.
Denote the corresponding functional of process $y(t)$ as $W(t)=\int_0^t U(y(t'))dt'$. Then the forward Feynman-Kac equation of the joint PDF $G(y,W,t)$ in Fourier space ($y\rightarrow k,W\rightarrow p$) is
\begin{equation}\label{FFKE_sub}
\begin{split}
    &\frac{\partial G(k,p,t)}{\partial t}= \mathcal{F}_y\{\phi_0(kg(y,t)) \mathcal{D}_t^{1-\alpha} G(y,p,t) \}  \\
    &~~-\mathcal{F}_y\left\{\frac{\partial}{\partial y}f(y,t) \mathcal{D}_t^{1-\alpha} G(y,p,t) + ipU(y)G(y,p,t) \right\},
\end{split}
\end{equation}
which recovers \eqref{FFKE_k} when $\alpha=1$; the detailed derivation is presented in Appendix \ref{AppB}. The symbol $\mathcal{D}_t^{1-\alpha}$ is the fractional substantial derivative operator introduced in \cite{FriedrichJenkoBauleEule:2006,SokolovMetzler:2003} with
\begin{equation*}
\begin{split}
    &\mathcal{D}_t^{1-\alpha} G(y,p,t)   \\
    &= \frac{1}{\Gamma(\alpha)}\left[\frac{\partial}{\partial t}+ipU(y)\right] \int_0^t\frac{e^{-(t-t')ipU(y)}}{(t-t')^{1-\alpha}}G(y,p,t')dt'.
\end{split}
\end{equation*}

\subsection{Special/particular cases}
This subsection provides some special/particular cases of the  derived equations in above subsection.

\begin{enumerate}
  \item {\it Generalized Fokker-Planck equation.} Let $p=0$ in \eqref{FFKE_k}. In this case, $G(x,p=0,t)=\int_0^{\infty}G(x,A,t)dA$ reduces to $G(x,t)$, the marginal PDF of finding the particle at position $x$ at time $t$. Correspondingly, the forward Feynman-Kac equation \eqref{FFKE_k} reduces to the generalized Fokker-Planck equation \cite{DenisovHorsthemkeHanggi:2009}, where three kinds of noises (Gaussian white noise, Poisson white noise and L\'{e}vy stable noise) are considered for the specific forms of this equation.
  \item {\it Gaussian white noise.} If the noise $\xi(t)$ is the Gaussian white noise in \eqref{FFKE_sub}, for arbitrary $f(x,t)$ and $g(x,t)$, we get the forward Feynman-Kac equation:
        \begin{equation}\label{FFKE_B3}
            \begin{split}
                \frac{\partial G(y,p,t)}{\partial t}=& \left[-\frac{\partial}{\partial y}f(y,t)+\frac{\partial^2}{\partial y^2} g^2(y,t)\right] \\
                 &\cdot\mathcal{D}_t^{1-\alpha}G(y,p,t)-ipU(y)G(y,p,t).
            \end{split}
        \end{equation}
        This equation is consistent with the forward Feynman-Kac equation with inverse $\alpha$-stable subordinator proposed in \cite{CairoliBaule:2017} by Langevin-type approach. Especially when $g(x,t)\equiv1$, it recovers the equation in \cite{CarmiBarkai:2011} by CTRW models.
  \item {\it Non-Gaussian $\beta$-stable noise.} If the noise $\xi(t)$ is the non-Gaussian $\beta$-stable noise in \eqref{FFKE_sub}, for arbitrary $f(x,t)$ and $g(x,t)$, the forward Feynman-Kac equation becomes
        \begin{equation}\label{FFKE_B4}
          \begin{split}
            \frac{\partial G(y,p,t)}{\partial t}=& \left[-\frac{\partial}{\partial y}f(y,t)+\nabla_y^\beta |g(y,t)|^\beta\right] \\
             & \cdot\mathcal{D}_t^{1-\alpha}G(y,p,t)-ipU(y)G(y,p,t),
          \end{split}
        \end{equation}
        where $\nabla_y^\beta$ is the Riesz space fractional derivative operator with Fourier symbol $-|k|^\beta$ \cite{WuDengBarkai:2016,CarmiTurgemanBarkai:2010}; and in $y$ space,
        \begin{equation*}
          \nabla_y^\beta h(y)=-\frac{{}_{-\infty}D_y^\beta h(y)+{}_yD_{\infty}^\beta h(y)}{2\cos(\beta\pi/2)},
        \end{equation*}
        where for $n-1<\beta<n$,
        \begin{equation*}
          {}_{-\infty}D_y^\beta h(y)=\frac{1}{\Gamma(n-\beta)}\frac{d^n}{dy^n}\int_{-\infty}^y \frac{h(y')}{(y-y')^{\beta+1-n}}dy',
        \end{equation*}
        \begin{equation*}
          {}_yD_{\infty}^\beta h(y)=\frac{(-1)^n}{\Gamma(n-\beta)}\frac{d^n}{dy^n}\int_y^{\infty} \frac{h(y')}{(y'-y)^{\beta+1-n}}dy'.
        \end{equation*}
        This equation extends \eqref{FFKE_B3} to L\'{e}vy stable noise, denoting the heavy-tailed jump length in CTRW models, which will be further studied by an application in the next section.
  \item {\it A positive functional.} If the functional $A$ is positive at any time $t$, the Fourier transform $A\rightarrow p$ will be replaced by the Laplace transform $G(x,p,t)=\int_0^{\infty}e^{-pA}G(x,A,t)dA$. Eventually, the forward Feynman-Kac equation corresponding to \eqref{FFKE_sub} is obtained by replacing $ip$ with $p$.
\end{enumerate}

\subsection{Backward equation}

The forward Feynman-Kac equation \eqref{FFKE_sub} describes the joint PDF $G(x,A,t)$ of position $x$ and functional $A$. But sometimes, especially in practical applications \cite{Majumdar:2005,CarmiTurgemanBarkai:2010,CarmiBarkai:2011}, what we are interested in may be only the distribution of functional $A$, which prompts us to develop the backward Feynman-Kac equation governing $G_{x_0}(A,t)$---the PDF of functional $A$ at time $t$, given that the process has started at $x_0$. In this subsection, the stochastic process we consider is 
\begin{equation}\label{LEmodelBe}
\dot{x}(t)=f(x(t))+g(x(t))\xi(t), 
\end{equation}
where $\xi(t)$ again is a L\'{e}vy noise.

Noting that $x_0$ here is a deterministic variable instead of a  random one, we should dig out how functional $A$ depends on initial position $x_0$. Different from the increment $\delta A$ considered in the forward Feynman-Kac equation, here we should build the relation between $A$ and $x_0$ as, during the time interval $\tau\,(\tau\rightarrow0)$,
\begin{equation}\label{SplitA}
  \begin{split}
    A(t+\tau)|_{x_0}&=\int_0^{\tau} U(x(t'))dt' + \int_\tau^{t+\tau}U(x(t'))dt' \\
    &=U(x_0)\tau+A(t)|_{x(\tau)},
  \end{split}
\end{equation}
where $A(t+\tau)|_{x_0}$ denotes the functional $A$ at time $t+\tau$ with the initial position $x_0$. Letting $t=0$ in \eqref{LEIto} gives the expression of $x(\tau)$:
\begin{equation}\label{xtau}
  x(\tau)=x_0+f(x_0)\tau+g(x_0)\eta(\tau).
\end{equation}
Expressing $G_{x_0}(A,t)$ in the Fourier space as
\begin{equation*}
  G_{x_0}(p,t)=\langle e^{-ipA(t)|_{x_0}}\rangle,
\end{equation*}
we could get the form of $G_{x_0}(p,t+\tau)$ from \eqref{SplitA} as:
\begin{equation}\label{Gpttau}
  G_{x_0}(p,t+\tau)=\langle \langle e^{-ipA(t)|_{x(\tau)}}\rangle\rangle e^{-ipU(x_0)\tau}.
\end{equation}
Since $A(t)|_{x(\tau)}$ denotes the functional $A$ at time $t$ with the initial position $x(\tau)$, it is independent of the event before $x(\tau)$, e.g., $\eta(\tau)$. So the internal angle bracket in \eqref{Gpttau} denotes the average of $A(t)|_{x(\tau)}$ while the external one the average of $\eta(\tau)$.
Then the increment $\delta G_{x_0}(p,t)$ can be expressed as
\begin{equation*}
\begin{split}
    \delta &G_{x_0}(p,t):=G_{x_0}(p,t+\tau)-G_{x_0}(p,t) \\
    &=\langle \langle e^{-ipA(t)|_{x(\tau)}}\rangle\rangle e^{-ipU(x_0)\tau}-\langle e^{-ipA(t)|_{x_0}}\rangle.
\end{split}
\end{equation*}
Taking $\tau\rightarrow0$, omitting the higher order term of $\tau$, we get
\begin{equation}\label{BFKE_k1}
\begin{split}
    \delta G_{x_0}(p,t)=&~\langle \langle e^{-ipA(t)|_{x(\tau)}}\rangle\rangle-\langle e^{-ipA(t)|_{x_0}}\rangle  \\
    &-ipU(x_0)\tau \langle e^{-ipA(t)|_{x_0}}\rangle,
\end{split}
\end{equation}
where the last term equals to $-ipU(x_0)\tau G_{x_0}(p,t)$.
Next, we will deal with the first two terms in the right hand side of \eqref{BFKE_k1} carefully by keeping the terms containing $\mathcal{O}(\tau)$ but removing the terms $o(\tau)$.

Taking Fourier transform $x_0\rightarrow k_0$ in \eqref{BFKE_k1}, then $\langle e^{-ipA(t)|_{x_0}}\rangle$ becomes $G_{k_0}(p,t)$. But for $\langle \langle e^{-ipA(t)|_{x(\tau)}}\rangle\rangle$, it is not easy to get the form in Fourier space. In this part, we take $g(x)\equiv1$, i.e., the noise in this system is additive noise. The case that function $g(x)$ depends on position $x$ will be considered in Appendix \ref{AppA}.

For convenient, we denote $T_\eta=\langle e^{-ipA(t)|_{x(\tau)}}\rangle$. Since $g(x)\equiv1$, \eqref{xtau} becomes $x(\tau)=x_0+f(x_0)\tau+\eta(\tau)$, where $f(x_0)$ depends on the initial position $x_0$. Therefore, $x(\tau)$ is not a simple shift of $x_0$ and we write the Fourier transform ($x_0\rightarrow k_0$) of $\langle T_\eta\rangle$ as
\begin{equation*}
  \mathcal{F}_{x_0}\{\langle T_\eta \rangle\}=\left\langle \int_{-\infty}^{\infty}e^{-ik_0x(\tau)}T_\eta e^{ik_0(f(x_0)\tau+\eta(\tau))}dx_0  \right\rangle.
\end{equation*}
Then we turn $dx_0$ into $dx(\tau)$ and get
\begin{equation}\label{FTeta0}
  \begin{split}
    \mathcal{F}_{x_0}\{\langle T_\eta \rangle\}
    =&\Big\langle \int_{-\infty}^{\infty}e^{-ik_0x(\tau)}T_\eta e^{ik_0(f(x_0)\tau+\eta(\tau))}dx(\tau)  \Big\rangle  \\
    -\Big\langle \int_{-\infty}^{\infty}&e^{-ik_0x(\tau)}T_\eta e^{ik_0(f(x_0)\tau+\eta(\tau))}\frac{d f(x_0)}{d x_0}\tau dx_0  \Big\rangle.
  \end{split}
\end{equation}
Since all $x_0$ and $f(x_0)$ are multiplied by $\tau$ in \eqref{FTeta0}, replacing all $x_0$ by $x(\tau)$ in \eqref{FTeta0} yields higher-order terms of $\tau$, which can be omitted. Then writing $e^{ik_0f(x_0)\tau}\simeq1+ik_0f(x_0)\tau$, the first term on the right hand side of \eqref{FTeta0} reduces to
\begin{equation*}
  \begin{split}
    \Big\langle &\int_{-\infty}^{\infty}e^{-ik_0x(\tau)}T_\eta e^{ik_0\eta(\tau)}dx(\tau)  \Big\rangle  \\
    &+ik_0\tau \Big\langle \int_{-\infty}^{\infty}e^{-ik_0x(\tau)}T_\eta f(x(\tau))dx(\tau)  \Big\rangle,
  \end{split}
\end{equation*}
where the latter term of above equals to
\begin{equation}\label{ReverseOperator}
  \tau\mathcal{F}_{x_0}\Big\{\frac{\partial}{\partial x_0}f(x_0)G_{x_0}(p,t)\Big\}.
\end{equation}
The second term on the right hand side of \eqref{FTeta0} gives
\begin{equation*}
\begin{split}
   - \tau \Big\langle \int_{-\infty}^{\infty} &e^{-ik_0x(\tau)}T_\eta \frac{d f(x(\tau))}{dx(\tau)}dx(\tau)\Big\rangle   \\
       &=-\tau\mathcal{F}_{x_0}\Big\{\frac{d f(x_0)}{d x_0}G_{x_0}(p,t)\Big\}.
\end{split}
\end{equation*}
Therefore, the Fourier transform of $\langle \langle e^{-ipA(t)|_{x(\tau)}}\rangle\rangle-\langle e^{-ipA(t)|_{x_0}}\rangle$ in \eqref{BFKE_k1}, replacing $x(\tau)$ by $y$, reduces to
\begin{equation*}
    \Big\langle \int_{-\infty}^{\infty}e^{-ik_0y}T_\eta (e^{ik_0\eta(\tau)}-1)dy  \Big\rangle
    +\tau\mathcal{F}\Big\{f(x_0)\frac{\partial G_{x_0}(p,t)}{\partial x_0}\Big\},
\end{equation*}
i.e.,
\begin{equation*}
  \tau\phi_0(-k_0)G_{k_0}(p,t)+\tau\mathcal{F}_{x_0}\Big\{f(x_0)\frac{\partial G_{x_0}(p,t)}{\partial x_0}\Big\}
\end{equation*}
on account of \eqref{Forward1}. Dividing \eqref{BFKE_k1} by $\tau$ and taking the limit $\tau\rightarrow0$, we obtain the backward Feynman-Kac equation in Fourier space:
\begin{equation}
\begin{split}
    &\frac{\partial G_{k_0}(p,t)}{\partial t}=\phi_0(-k_0)G_{k_0}(p,t) \\
    &~~~~+\mathcal{F}_{x_0}\Big\{f(x_0)\frac{\partial G_{x_0}(p,t)}{\partial x_0}-ipU(x_0)G_{x_0}(p,t)\Big\}.
\end{split}
\end{equation}

If the noise $\xi(t)$ is Gaussian white noise, then $\phi_0(-k_0)=-k_0^2$ and we get the backward Feynman-Kac equation:
\begin{equation}\label{BFKE_fG}
    \begin{split}
        \frac{\partial G_{x_0}(p,t)}{\partial t}&= \frac{\partial^2}{\partial x_0^2} G_{x_0}(p,t) \\
         &+ f(x_0)\frac{\partial}{\partial x_0}G_{x_0}(p,t)-ipU(x_0)G_{x_0}(p,t),
    \end{split}
\end{equation}
which is the same as the backward Feynman-Kac equation with $\alpha=1$ proposed in \cite{CarmiBarkai:2011} from CTRW models,
 $\alpha$ being the exponent characterizing the waiting time PDF in CTRW models or the subordinator PDF in Langevin system.

 If the noise $\xi(t)$ is non-Gaussian $\beta$-stable noise, i.e., $\phi_0(-k_0)=-|k_0|^\beta$, then the backward Feynman-Kac equation becomes
\begin{equation}\label{BFKE_fbeta}
  \begin{split}
    \frac{\partial G_{x_0}(p,t)}{\partial t}&= \nabla_{x_0}^\beta G_{x_0}(p,t) \\
     &+f(x_0)\frac{\partial}{\partial x_0}G_{x_0}(p,t)-ipU(x_0)G_{x_0}(p,t),
  \end{split}
\end{equation}
which is an extension for the backward Feynman-Kac equation in \cite{CarmiTurgemanBarkai:2010} based on CTRW models, in which  jump length obeys heavy-tailed distribution but without a force field $f(x)$. In the case that $g(x)$ is not a constant, we assume $\xi(t)$ to be Gaussian white noise and derive the backward Feynman-Kac equation as
\begin{equation}\label{BFKE}
\begin{split}
    \frac{\partial G_{x_0}(p,t)}{\partial t}&= g^2(x_0)\frac{\partial^2}{\partial x_0^2} G_{x_0}(p,t) \\
            &~+ f(x_0)\frac{\partial }{\partial x_0} G_{x_0}(p,t) - ipU(x_0)G_{x_0}(p,t),
\end{split}
\end{equation}
which goes back to \eqref{BFKE_fG} when $g(x_0)\equiv1$. See the detailed derivation in Appendix \ref{AppA}.

\section{Applications}\label{Sec3}

For the stochastic dynamics driven by additive white noise (or Gaussian jump length in CTRW models), there are a great quantify of applications for their corresponding Feynman-Kac equations \cite{CarmiBarkai:2011,CarmiTurgemanBarkai:2010}. Here we provide the applications for Feynman-Kac equations of more general stochastic processes discussed above. More concretely, two applications of the generalized Feynman-Kac equations are given, including the occupation time in the positive half of a particle moving in a box with multiplicative Gaussian white noise and the area under the curve of trajectory of the stochastic process with a quadratic potential driven by additive L\'{e}vy noise.


\subsection{Occupation time in the positive half of a box}
We first discuss the occupation time in $x>0$ for a particle moving freely but with a multiplicative Gaussian white noise in a box $[-L,L],\,L>0$, then give its direct application---the first-passage time.

\subsubsection{Distribution of occupation time}
We take $U(x_0)$ in \eqref{BFKE} to be $\Theta(x_0)$ ($\Theta(x)=1$ for $x\geq0$ and $\Theta(x)=0$ otherwise), and then get occupation time of a particle in the positive half-space as $T_+(t)=\int_0^t \Theta[x(t')]dt'$.
In this case, $T_+(t)$ is always positive. We replace the Fourier transform by Laplace transform in \eqref{BFKE} and remove $i$ in it.
To find the distribution of $T_+(t)$, we take the Laplace transform of the backward Feynman-Kac equation \eqref{BFKE} ($t\rightarrow s$):
\begin{equation}\label{App1}
\begin{split}
    sG_{x_0}(p,s)-1=&~g^2(x_0)\frac{\partial^2 }{\partial x_0^2}G_{x_0}(p,s)  \\
    &+f(x_0)\frac{\partial }{\partial x_0} G_{x_0}(p,s)-pU(x_0)G_{x_0}(p,s).
\end{split}
\end{equation}

\begin{figure}
  \centering
  \includegraphics[scale=0.5]{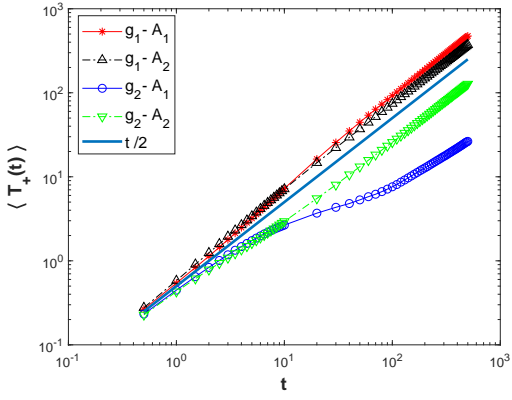}\\
  \caption{(Color online) Mean value of the occupation time $T_+$ in positive half-space for a particle moving in the box $[-1,1]$. Here $g_1$ represents the case $g(x)=AL-x$ and $g_2$ describes the case $g(x)=AL+x$. The other parameters are $A_1=1.1$ and $A_2=2$. Four kinds of different cases ($g_1,A_1$ with star-markers; $g_1,A_2$ with triangle markers; $g_2,A_2$ with inverted-triangle markers; $g_2,A_1$ with circle markers) are simulated with $1000$ trajectories and the total time $T=500$.}\label{fig1}
\end{figure}

\begin{figure}
  \centering
  \includegraphics[scale=0.5]{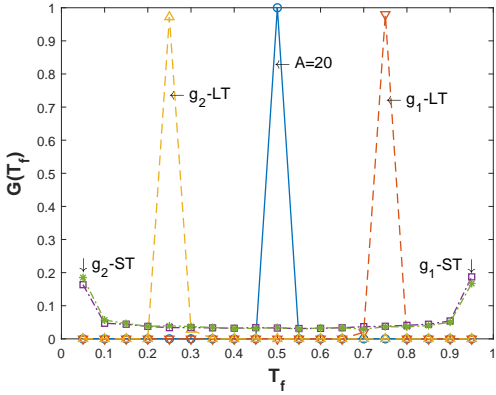}\\
  \caption{(Color online) PDF of the occupation fraction $T_f$ in positive half-space for a particle moving in the box $[-1,1]$. Here the PDF of $T_f$ for long times ($T=500$) and short times ($T=0.01$) are shown together, recorded as ``LT'' and ``ST'', respectively. 1000 trajectories are used. In this figure, $g_1$ and $g_2$ represent the cases $g(x)=AL-x$ and $g(x)=AL+x$, respectively. The solid line denotes $G(T_f)$ in long times with $A=20$ as well as $g=AL\pm x$ (the lines coincide for the cases $g=g_1$ and $g=g_2$).
  	Except the solid line, the other lines represent the case $A=2$, but $g$ or times are different.}\label{fig2}
\end{figure}

Here we pay attention to the effect of multiplicative noise, so the special choice of $f(x_0)=0$ and $g(x_0)=AL-x_0$ with $A>1$ is considered; and 
we also take $g(x_0)=AL+x_0$ to examine the effect of different sign of multiplicative noise. Interestingly/surprisingly, the theoretical results are quite close, just replacing $A+1$ by $A-1$ in \eqref{T+t1} and \eqref{GTf1}. The simulation results for the cases $g(x_0)=AL\pm x_0$ are shown together in Figs. \ref{fig1} and \ref{fig2}.

For the case of $g(x_0)=AL-x_0$, \eqref{App1} becomes
\begin{equation*}
  (AL-x_0)^2\frac{\partial^2 G_{x_0}(p,s)}{\partial x_0^2}-(s+pU(x_0))G_{x_0}(p,s)=-1.
\end{equation*}
By a variable substitution $y=AL-x_0>0$, the celebrated Euler equation is obtained:
\begin{equation*}
  y^2\frac{\partial^2 \tilde{G}_{y}(p,s)}{\partial y^2}-(s+p\tilde{U}(y))\tilde{G}_{y}(p,s)=-1.
\end{equation*}
It can be solved by a new variable substitution $y=e^t$.
Finally, we get the solutions of \eqref{App1} in two half-spaces, respectively,
\begin{equation}\label{App_Gx0}
\begin{split}
  G&_{x_0}(p,s)  \\
&=\left\{
  \begin{array}{ll}
    C_1(AL-x_0)^{\lambda_1}+C_2(AL-x_0)^{\lambda_2}+\frac{1}{s+p}  &x_0>0  \\
    C_3(AL-x_0)^{\lambda_3}+C_4(AL-x_0)^{\lambda_4}+\frac{1}{s}    &x_0<0,
  \end{array}\right.
\end{split}
\end{equation}
where
\begin{equation}\label{Fourroot}
  \lambda_{1,2}=\frac{1\mp\sqrt{1+4(s+p)}}{2}, \quad \lambda_{3,4}=\frac{1\mp\sqrt{1+4s}}{2}.
\end{equation}
Specify the reflecting boundary condition to \eqref{App_Gx0}, i.e.,  
\begin{equation}\label{BCref}
  \left.\frac{\partial G_{x_0}(p,s)}{\partial x_0}\right|_{x_0=\pm L}=0.
\end{equation}
The two conditions of \eqref{BCref} together with another two conditions ($G_{x_0}(p,s)$ and its derivative are continuous at $x_0=0$) can solve the four coefficients $C_{1-4}$ in \eqref{App_Gx0}. Then we get the final solution $G_{x_0}(p,s)$ at $x_0=0$:
\begin{equation}\label{Gps}
  G_0(p,s)=\frac{p}{s(p+s)}\cdot \frac{F_1F_2}{F_3F_4-F_1F_2}+\frac{1}{s},
\end{equation}
where
\begin{equation*}
  \begin{split}
    &F_1=A^{\lambda_4}-\frac{\lambda_4}{\lambda_3}(A+1)^{\lambda_4-\lambda_3}A^{\lambda_3},  \\
    &F_2=\lambda_2[A^{\lambda_2}-(A-1)^{\lambda_2-\lambda_1}A^{\lambda_1}],   \\
    &F_3=\lambda_4[A^{\lambda_4}-(A+1)^{\lambda_4-\lambda_3}A^{\lambda_3}],   \\
    &F_4=A^{\lambda_2}-\frac{\lambda_2}{\lambda_1}(A-1)^{\lambda_2-\lambda_1}A^{\lambda_1}.
  \end{split}
\end{equation*}

Equation \eqref{Gps} is the PDF of $T_+$ in Laplace space, but it cannot be inverted easily.
As usual, if one is concerned about the first moment of the occupation time $T_+(t)$, it can be computed by taking an inverse Laplace transform \cite{KlafterSokolov:2011} of
\begin{equation*}
  \langle T_+(s)\rangle= -\left.\frac{\partial G_0(p,s)}{\partial p}\right|_{p=0}.
\end{equation*}
By this formula, from \eqref{Gps} one can get 
\begin{equation}\label{T+s}
  \langle T_+(s)\rangle= -\frac{1}{s^2}\cdot \left.\frac{F_1F_2}{F_3F_4-F_1F_2}\right|_{p=0}.
\end{equation}
For long times, i.e., $s\ll1$, $(\lambda_1=\lambda_3\sim -s,\lambda_2=\lambda_4\sim 1)$,
\begin{equation}\label{T+t1}
  \langle T_+(t) \rangle\simeq\frac{A+1}{2A}t.
\end{equation}
While for short times, i.e., $s\gg1$, $(\lambda_1=\lambda_3\sim -\sqrt{s},\lambda_2=\lambda_4\sim\sqrt{s})$,
\begin{equation}\label{T+t2}
  \langle T_+(t) \rangle\simeq\frac{1}{2}t.
\end{equation}

It can be seen that for both long times and short times, $\langle T_+(t)\rangle$ scales asymptotically as $t$, which is also verified in Fig. \ref{fig1}. Four curves begin as $t/2$ and finally turn to $\frac{A+1}{2A}t$ for the case $g(x)=AL-x$ or $\frac{A-1}{2A}t$ for the case $g(x)=AL+x$.
Therefore, it is natural to consider the PDF of the occupation fraction $T_f\equiv T_+/t$.

For long times, i.e., $s\ll1$, together with $p\ll1$ due to the scale of $T_+(t)$,  we have $\lambda_1\sim -(s+p),\lambda_2\sim 1,\lambda_3\sim -s,\lambda_4\sim 1$ from \eqref{Fourroot} and $F_1\sim (A+1)/s$, $F_2\sim1$, $F_3\sim-1$, $F_4\sim (A-1)/(s+p)$, which gives the asymptotic expression of \eqref{Gps}:
\begin{equation*}
  G_0(p,s)\simeq\frac{2A}{2As+(A+1)p}.
\end{equation*}
By the inversion of the scaling form of a double Laplace transform in \cite{Godreche:2001}, after some calculations, using the nascent delta function:
\begin{equation*}
  \underset{\epsilon\rightarrow0}{\lim}\frac{\epsilon}{\pi(x^2+\epsilon^2)}=\delta(x),
\end{equation*}
we obtain the PDF of $T_f$:
\begin{equation}\label{GTf1}
  G(T_f)\simeq\frac{r}{T_f}\cdot \delta(T_f-r)\overset{d}{=}\delta(T_f-r),
\end{equation}
where $r=\frac{A+1}{2A}$ and $\overset{d}{=}$ denotes identical distribution.
Note that the PDF of $T_f$ in \eqref{GTf1} is normalized. Especially, $T_f$ reduces to a deterministic event for large $t$, occurring at $r$ with probability $1$.
But the value $r$ depends on $A$. When $A$ is sufficiently  large, this value will approach $\frac{1}{2}$ (see the curve for $A=20$ which has a peak at $\frac{1}{2}$ in Fig. \ref{fig2}). This phenomenon has an intuitive explanation that in this case the multiplicative noise term approximates an additive noise term $AL$ and thus it is consistent with the case of $\alpha=1$ in \cite{CarmiBarkai:2011}. On the other hand, when $A$ is small and close to $1$, the value $r$ is near $1$, which means that the particle stays in positive half-plane all the time. This phenomenon results from the multiplicative noise term. We simulate $G(T_f)$ with $A=2$ and it has a peak at $\frac{A+1}{2A}$ for $g(x)=AL-x$ (see $g_1-$ $\mathrm{LT}$ in Fig. \ref{fig2}) and a peak at $\frac{A-1}{2A}$ for $g(x)=AL+x$ (see $g_2-$ $\mathrm{LT}$ in Fig. \ref{fig2}).

For short times, i.e., $s\gg1$, we have $\lambda_1\sim-\sqrt{s+p},\,\lambda_2\sim\sqrt{s+p},\,\lambda_3\sim-\sqrt{s},\,\lambda_4\sim\sqrt{s}$ from \eqref{Fourroot} and $F_1\sim(A+1)^{2\sqrt{s}}A^{-\sqrt{s}}$, $F_2\sim\sqrt{s+p}A^{\sqrt{s+p}}$, $F_3\sim-\sqrt{s}(A+1)^{2\sqrt{s}}A^{-\sqrt{s}}$, $F_4\sim A^{\sqrt{s+p}}$, which result in the asymptotic expression of \eqref{Gps}:
\begin{equation*}
  G_0(p,s)\simeq-\frac{p}{s(p+s)}\cdot \frac{\sqrt{s+p}}{\sqrt{s}+\sqrt{s+p}}+\frac{1}{s}.
\end{equation*}
Then we obtain the PDF of $T_f$:
\begin{equation}\label{GTf2}
  G(T_f)\simeq\frac{1}{\pi}\cdot \frac{1}{\sqrt{x}\sqrt{1-x}},
\end{equation}
which is consistent with the classical Brownian functional \cite{Majumdar:2005}.
This result is as expected since for short times the particle does not interact with the boundaries and behaves like a free particle. Furthermore, if the time $t$ is sufficiently small, such that $x\ll AL$, then the multiplicative noise term approximates an additive noise term $AL$, so the PDFs of occupation fractions $T_f$ in cases $g(x)=AL\pm x$ all become the Lamperti PDF and present a symmetric curve with two peaks at $T_f=0$ and $T_f=1$ (see $g_2-$ $\mathrm{ST}$ and $g_1-$ $\mathrm{ST}$ in Fig. \ref{fig2}). Though $x\ll AL$, there is still a slight difference between two kinds of the multiplicative noises $g(x)=AL\pm x$. Therefore, the two curves in Fig. \ref{fig2} look a little skew to one side ($0$ or $1$).

For both long times and short times, in another perspective, the particle driven by the multiplicative noise term $g(x)=AL-x$ is more likely to move to the positive half-space since the distribution of $T_f$ has a larger proportion on the right side of $0.5$ in Fig. \ref{fig2}. On the contrary, for $g(x)=AL+x$, the distribution of $T_f$ concentrates on the left side of $0.5$.
This phenomenon can be explained by the corresponding Fokker-Plank equation of \eqref{FFKE_B3}. Taking $\alpha=1,\,p=0,\,f(x,t)=0$ in \eqref{FFKE_B3} and using the notation $g'(x)=dg(x)/dx$  give
\begin{equation}
    \begin{split}
        \frac{\partial G(x,t)}{\partial t}
        &= \frac{\partial^2}{\partial x^2} g^2(x)G(x,t) \\
        &= g^2(x)\frac{\partial^2G(x,t)}{\partial x^2}+4g(x)g'(x)\frac{\partial G(x,t)}{\partial x}   \\
         &~~~ +2\Big(g'(x)^2+g(x)g''(x)\Big)G(x,t),
    \end{split}
\end{equation}
where the coefficient $E:=4g(x)g'(x)$ in front of the first derivative of $G(x,t)$ is called noise induced drift \cite{CoffeyKalmykovWaldron:2004}. The cases of $g(x)=AL-x$ and $g(x)=AL+x$ lead to $E=4(x-AL)<0$ and $E=4(x+AL)>0$,  respectively, which means that the multiplicative noise term $g(x)=AL-x$ induces a positive drift while $g(x)=AL+x$ a negative drift.

\subsubsection{Distribution of first-passage time}
\begin{figure}
  \centering
  \includegraphics[scale=0.5]{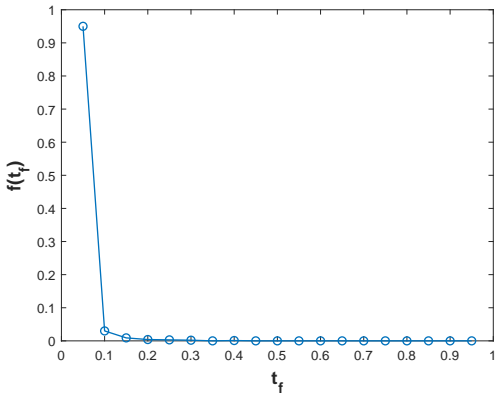}\\
  \caption{PDF of the first passage time of a particle in the box $[-1,1]$ starting at $-b$ and reaching $x=0$ for the first time. We simulate it with $1000$ trajectories and the total time $T=10$. The parameters are $b=1/2,\,A=2$, and then $C_{Ab}=0.0565$, which is consistent with the curve that has a peak near $0.0565$ in the figure.}\label{fig3}
\end{figure}

The application of occupation time in the previous part is a good beginning to consider a problem of the first-passage time $t_f$. Still assuming a particle moves freely in the box $[-L,L]$, $t_f$ denotes the time it takes a particle starting at $x_0=-bL,\,0<b<1$ to reach $x=0$ for the first time \cite{Redner:2001}. The distribution of $t_f$ can be obtained from the occupation time functional by using an identity due to Kac \cite{Kac:1951}:
\begin{equation*}
  \mathbb{P}(t_f>t)=\mathbb{P}\Big(\underset{0\leq\tau\leq t}{\max}x(\tau)<0\Big)=\underset{p\rightarrow\infty}{\lim}G_{x_0}(p,t),
\end{equation*}
where $G_{x_0}(p,t)$ is the Laplace transform of the PDF of functional $T_+$ in the previous subsection.
If the particle has crossed $x=0$ at time $t$, we have $T_+>0$ and $e^{-pT_+}=0$ for $p\rightarrow\infty$. Then two sides of the last equation equal to $0$. Otherwise $T_+=0$ and $e^{-pT_+}=1$ lead two sides equal $1$.
So now, taking $x_0=-bL$ in \eqref{App_Gx0} in the previous subsection, we get
\begin{equation}\label{G-bL}
  G_{-bL}(p,s)=\frac{p}{s(p+s)}\cdot \frac{F_{1b}F_2}{F_3F_4-F_1F_2}+\frac{1}{s},
\end{equation}
where $F_{1-4}$ are the same as the ones in \eqref{Gps} and
\begin{equation*}
  F_{1b}=(A+b)^{\lambda_4}-\frac{\lambda_4}{\lambda_3}(A+1)^{\lambda_4-\lambda_3}(A+b)^{\lambda_3}.
\end{equation*}
When $p\rightarrow\infty$, we consider the long-time behaviour (i.e., $s\rightarrow0$) and have $\lambda_1\sim-\sqrt{p},\lambda_2\sim\sqrt{p},\lambda_3\sim-s,\lambda_4\sim1$. Substituting $\lambda_{1-4}$ into \eqref{G-bL} yields
\begin{equation*}
  \underset{p\rightarrow\infty}{\lim}G_{-bL}(p,s)\simeq\ln\left(1+\frac{b}{A}\right)-\frac{b}{1+A}=:C_{Ab},
\end{equation*}
which is a constant only depending on $A$ and $b$.
Considering the first-passage time PDF satisfying $f(t)=\frac{\partial}{\partial t}[1-\mathbb{P}(t_f>t)]$, we have the PDF of $t_f$ in Laplace $s$ space
\begin{equation*}
  f(s)\simeq 1-C_{Ab}s\simeq e^{-C_{Ab}s},
\end{equation*}
and thus
\begin{equation*}
  f(t_f)\simeq \delta(t_f-C_{Ab}).
\end{equation*}
This means that the first-passage time is a deterministic event, occurring at $C_{Ab}$ with probability $1$; see the distribution of first passage time $t_f$ in Fig. \ref{fig3}. Furthermore, for $0<b<1<A$, $C_{Ab}$ is monotonously increasing of $b$ but decreasing of $A$, being the same as physical intuition.

\subsection{Area under the random walk curve}
Now we turn to an application of the Langevin system containing a force field and non-Gaussian $\beta$-stable noise. In this case, we take $U(x)=x$ and get the functional $A_x=\int_0^t x(t')dt'$, denoting the total area under the curve of trajectory $x(t)$ \cite{BauleFriedrich:2006,Grebenkov:2007}. This functional $A_x$ is also related to the phase accumulated by spins in an NMR experiment \cite{Grebenkov:2007}. Since the analytical solutions of $G_{x_0}(p,t)$ in \eqref{BFKE_fbeta} cannot be easily obtained due to the Riesz space fractional derivative operator $\nabla_x^\beta$, we resort to the forward Feynman-Kac equation \eqref{FFKE_B4} by integrating the solution $G(x,p,t)$ over $x$ with initial position $x_0$ to get the marginal PDF of $G_{x_0}(p,t)$.

In the case of a quadratic potential, where $V(x,t)=bx^2/2\,(b>0)$, ($f(x,t)=-\partial V(x,t)/\partial x=-bx$) and $g(x,t)\equiv1$, $U(x)=x$, $\alpha=1$, the forward Feynman-Kac equation \eqref{FFKE_k} takes the form
\begin{equation*}
  \frac{\partial G(k,p,t)}{\partial t}+(bk-p)\frac{\partial }{\partial k}G(k,p,t)=\phi_0(k)G(k,p,t).
\end{equation*}
Its general solution is given by the following \cite{PolyaninZaitsevMoussiaux:2002}:
\begin{equation}\label{APP2_1}
\begin{split}
    G(k,p,t)=&~\exp\left[\int_0^k\frac{\phi_0(z)}{bz-p}dz+c_1\right] \\
    &~\cdot\Psi\left[\frac{1}{b}\ln|bk-p|-t+c_2\right],
\end{split}
\end{equation}
where $c_1,c_2$ are constants and $\Psi(x)$ is an arbitrary function. Using the initial condition $G(k,p,0)=1$ (the particle starts at $x_0=0$), we get
\begin{equation}\label{APP2_2}
    \Psi\left[\frac{1}{b}\ln|bk-p|+c_2\right]=\exp\left[-\int_0^k\frac{\phi_0(z)}{bz-p}dz-c_1\right].
\end{equation}
Then replacing $k$ by $l(k):=\frac{bk-p}{be^{bt}}+\frac{p}{b}$ in \eqref{APP2_2} yields
\begin{equation*}
  \Psi\left[\frac{1}{b}\ln|bk-p|-t+c_2\right]=\exp\left[-\int_0^{l(k)}\frac{\phi_0(z)}{bz-p}dz-c_1\right].
\end{equation*}
Substituting this result into \eqref{APP2_1}, we obtain
\begin{equation*}
  G(k,p,t)=\exp\left[\int_{l(k)}^k\frac{\phi_0(z)}{bz-p}dz\right].
\end{equation*}
Letting $k=0$, we get the PDF of functional $A_x$ in Fourier space $(A_x\rightarrow p)$:
\begin{equation}\label{APP2_Gpt}
  G(p,t):=G(k,p,t)|_{k=0}=\exp\left[\int_{\frac{p}{b}(1-e^{-bt})}^0 \frac{\phi_0(z)}{bz-p}dz\right].
\end{equation}

Now we discuss the specific dynamical behaviour of functional $A_x$ with L\'{e}vy $\beta$-stable noise ($\phi_0(k)=-|k|^\beta$). Considering a variable substitution $z=\frac{p}{b}(1-e^{-bt})y$ in \eqref{APP2_Gpt}, we obtain
\begin{equation}\label{APP2_Gpt2}
  \ln G(p,t)=-C_b(t)\left(\frac{1-e^{-bt}}{b}\right)^{\beta+1} |p|^\beta ,
\end{equation}
where $C_b(t)$ is independent of $p$ \cite{GradshteynRyzhikGeraniumsTseytlin:1980}:
\begin{equation*}
  \begin{split}
    C_b(t)&=\int_0^1\frac{y^\beta}{1-(1-e^{-bt})y}dy  \\
    &=B(\beta+1,1)\cdot {}_2F_1(1,\beta+1;\beta+2;1-e^{-bt}).
  \end{split}
\end{equation*}
It can be seen from \eqref{APP2_Gpt2} that the functional $A_x$ also obeys L\'{e}vy $\beta$-stable distribution. Next what we need to pay attention to is the coefficient in front of $|p|^\beta$ in \eqref{APP2_Gpt2}.

For long times $t\rightarrow\infty$, we find that
\begin{equation*}
  C_b(t)=\int_0^1\frac{y^\beta}{1-(1-e^{-bt})y}dy \simeq bt,
\end{equation*}
since this integral scales as $bt$ in both two extreme cases ($\beta=0$ and $\beta=2$). Substituting it into \eqref{APP2_Gpt2}, we get
\begin{equation}\label{APP2_Gpt3}
  G(p,t)\simeq \exp(-b^{-\beta}t|p|^\beta)~~~~~~\textrm{as}~~~ t\rightarrow\infty.
\end{equation}
For short times $t\rightarrow0$, $_2F_1(1,\beta+1;\beta+2;1-e^{-bt})\sim1$, and thus
\begin{equation}\label{APP2_Gpt4}
  G(p,t)\simeq \exp\left(-\frac{t^{\beta+1}}{\beta+1}|p|^\beta\right)  ~~~~~~\textrm{as}~~~ t\rightarrow0.
\end{equation}

\begin{figure}
  \centering
  \includegraphics[scale=0.5]{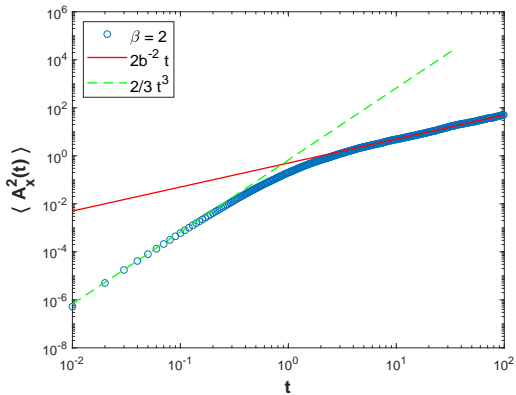}\\
  \caption{(Color online) Second moment $\langle A_x^2 \rangle$ of the area under the trajectory curve with $\beta=2$ and $b=2$, generated with $1000$ trajectories and the total time $T=100$. The circle-markers denote the simulation results. The dotted line denotes the theoretical result $\langle A_x^2 \rangle \simeq\frac{2}{3}t^3$ for short time while the solid line represents $\langle A_x^2 \rangle\simeq\frac{2}{b^2}t$ for long time. This figure shows a crossover of scaling regimes from $t^3$ to $t$.}\label{fig4}
\end{figure}

For the special case $\beta=2$, i.e., Gaussian white noise, by the formula
\begin{equation*}
  \langle A_x^2\rangle= \left.\frac{\partial^2}{\partial p^2}G(p,t)\right|_{p=0},
\end{equation*}
we get
\begin{equation}\label{Ax2LT}
  \langle A_x^2\rangle \simeq 2b^{-2}t,  ~~~~~~\textrm{as}~~~ t\rightarrow\infty,
\end{equation}
and
\begin{equation}\label{Ax2ST}
  \langle A_x^2\rangle \simeq \frac{2}{3}t^3,  ~~~~~~\textrm{as}~~~ t\rightarrow0,
\end{equation}
which are verified by numerical simulations. In Fig. \ref{fig4}, the functional $A_x$ exhibits a crossover between different scaling regimes (from $t^3$ to $t$).
When the particle begins its movement from the origin, i.e., $x\ll1$, the effect of force ($f=-bx$) can be omitted. As time goes on, this effect is getting bigger and bigger, and eventually produces the multi-scale phenomenon. On the contrary, for the case without the force field $f$, i.e., $b=0$,
it is equivalent to that $b\rightarrow0$ for any $t$ from \eqref{APP2_Gpt2}. Then only the single-scale phenomenon $\langle A_x^2\rangle \simeq \frac{2}{3}t^3$ can be observed, which is consistent with \cite{CarmiTurgemanBarkai:2010} by taking $\alpha=1$ there.

\begin{figure}
  \centering
  \includegraphics[scale=0.5]{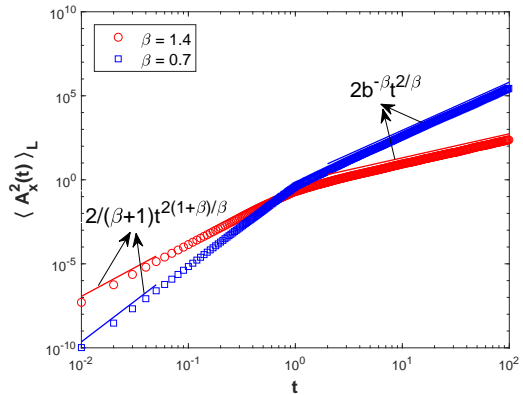}\\
  \caption{(Color online) Pseudo second moment $\langle A_x^2\rangle_L$ by a cut-off approach, generated with $1000$ trajectories and the total time $T=100$ with $\beta=1.4,\,0.7,$ and $b=2$. The circle-markers and square-markers denote the simulation results of $\beta=1.4$ and $\beta=0.7$,  respectively. The solid lines are the theoretical results with slope $2(\beta+1)/\beta$ for short time and $2/\beta$ for long time. It shows that for different L\'{e}vy $\beta$-stable noise ($0<\beta<2$), there is a crossover of scaling regimes from $t^{2(\beta+1)/\beta}$ to $t^{2/\beta}$.}\label{fig5}
\end{figure}

As for the general case $0<\beta<2$, the mean squared displacement of $A_x$ diverges \cite{MetzlerKlafter:2000}: $\langle A_x^2\rangle\rightarrow\infty$. But for a particle with non-diverging mass, a finite velocity of propagation exists, making long instantaneous jumps impossible.
Their fractional moments can be written as
\begin{equation}\label{seondmoment}
  \langle |A_x|^\delta\rangle\propto \tilde{t}^{\delta/\beta},
\end{equation}
where $0<\delta<\beta<2$. From \eqref{APP2_Gpt3} and \eqref{APP2_Gpt4}, one can get that in (\ref{seondmoment}) $\tilde{t}$ should be  $t^{\beta+1}$ for short times and $t$ for long times.
So we rescale the fractional moments and get the pseudo second moment $[A_x^2]\propto \tilde{t}^{2/\beta}$. An alternative method is to consider the $(A_x-t)$ scaling relations, or to measure the width of the PDF $G(A_x,t)$ rather than its variance \cite{MetzlerKlafter:2000}. More precisely, enclose the particle in an imaginary growing box \cite{JespersenMetzlerFogedby:1999} and define
\begin{equation*}
  \langle A_x^2\rangle_L := \int_{L_1t^{1/\beta}}^{L_2t^{1/\beta}}A_x^2G(A_x,t)dA_x \simeq \tilde{t}^{2/\beta},
\end{equation*}
where $L_1$ and $L_2$ are chosen to adapt the scaling regimes in \eqref{APP2_Gpt3} and \eqref{APP2_Gpt4}, i.e., for long time $-L_1=L_2=\sqrt{2b^{-\beta}}$ while for short time $-L_1=L_2=\sqrt{2/(1+\beta)}$.
This has been implemented numerically and can be seen in Fig. \ref{fig5}. We take $\beta$ to be $1.4$ or $0.7$ and $b=2$. The markers denote simulation results while the solid lines are the theoretical ones
\begin{equation*}
  \langle A_x^2\rangle \simeq 2b^{-\beta}t^{2/\beta},  ~~~~~~\textrm{as}~~~ t\rightarrow\infty,
\end{equation*}
and
\begin{equation*}
  \langle A_x^2\rangle \simeq \frac{2}{\beta+1}t^{2(\beta+1)/\beta},  ~~~~~~\textrm{as}~~~ t\rightarrow0,
\end{equation*}
which go back to \eqref{Ax2LT} and \eqref{Ax2ST}, respectively, when $\beta=2$.

\section{Summary and discussion}\label{Sec4}

The Feynman-Kac equations have striking advantages in characterizing the PDFs of various general statistical quantities.  Under the CTRW framework, there have been a systematic derivations of the equations. But the CTRW models can not well describe the multiplicative noise and the arbitrary additive noise together with force field, being more conveniently modeled by the Langevin system. 


The contributions of the paper are twofold: deriving the forward Feynman-Kac equation from the overdamped Langevin equation driven by an arbitrary L\'{e}vy noise together with a time-dependent multiplicative noise term and an arbitrary time-dependent external force field; and deriving the backward Feynman-Kac equation with an arbitrary additive L\'{e}vy noise or a multiplicative Gaussian white noise, together with an arbitrary force field.
For some special noises and force fields, the obtained equations are consistent with the existing works. Two applications of the derived equations to solve PDFs of the occupation time $T_+$ and the total area $A_x$ under the curve of the particle trajectory are carefully provided. In the first application, we take a multiplicative Gaussian white noise and restrict the particle in a box $[-L,L]$ with reflecting boundary conditions. Then we find a new phenomenon that for long times the PDF of occupation fraction is a $\delta$-function. In the second application, we take an additive L\'{e}vy $\beta$-stable noise and find that the $A_x$ also obeys L\'{e}vy stable distribution but experiences a crossover between different scaling regimes.

Using the techniques of subordinator in deriving Feynman-Kac equations presented in \cite{CairoliBaule:2017}, we also derive the forward Feynman-Kac equations from the coupled Langevin equation with $\alpha$-stable subordinator and arbitrary L\'{e}vy noise based on the Langevin framework.

\begin{acknowledgments}
This work was supported by the National Natural Science Foundation of China under grant no. 11671182, and the Fundamental Research Funds for the Central Universities under grants no. lzujbky-2018-ot03 and no. lzujbky-2017-ot10.
\end{acknowledgments}

\appendix
\section{Forward Feynman-Kac equation with subordinator}\label{AppB}

Since the forward Feynman-Kac equation in the case of Gaussian white noise $\xi(s)$ has been derived in \cite{CairoliBaule:2017}, we can make the best of some techniques of subordinator in that paper and extend its result to arbitrary L\'{e}vy noise $\xi(s)$ in the Langevin framework. 
Some of the calculations about subordinator may be omitted for simplicity; see \cite{CairoliBaule:2017} for the details.

Since $y(t)=x(S(t))$, we can build the Langevin equation of $y(t)$ from model \eqref{LEmodelB} as:
\begin{equation*}
  \dot{y}(t)=f(y(t),t)\dot{S}(t)+g(y(t),t)\xi(S(t))\dot{S}(t).
\end{equation*}
Being similar to \eqref{LEIto}, with the It\^{o} interpretation, the increment of $y(t)$ reads
\begin{equation}\label{LEItoB}
  \delta y(t)=f(y(t),t)\delta S(t) + g(y(t),t)\delta\eta(S(t)),
\end{equation}
where $\delta S(t)=S(t+\tau)-S(t)$ and $\delta\eta(S(t))=\eta(S(t+\tau))-\eta(S(t))$. Next, similar to \eqref{dGkpt2}, we obtain the increment of $G(y,W,t)$ in Fourier space $(y\rightarrow k,W\rightarrow p)$:
\begin{equation}\label{dGkptB}
\begin{split}
    \delta G(k,p,t)=&~ \langle e^{-iky(t)-ipW(t)}(e^{-ikg(y(t),t)\delta\eta(S(t))}-1) \rangle \\
    &-ik \langle e^{-iky(t)-ipW(t)}f(y(t),t)\delta S(t) \rangle \\
    &-ip\tau \langle e^{-iky(t)-ipW(t)}U(y(t)) \rangle,
\end{split}
\end{equation}
where the first term on the right hand side can be reduced to
\begin{equation*}
  \langle e^{-iky(t)-ipW(t)}\phi_0(kg(y(t),t)) \, \delta S(t) \rangle
\end{equation*}
as usual by the characteristic function of $\delta\eta(t)$ in \eqref{NoiseChar}.
So dividing \eqref{dGkptB} by $\tau$ and taking the limit $\tau\rightarrow0$, we obtain
\begin{equation}\label{dGkpt2B}
\begin{split}
    \frac{\partial}{\partial t} G(k,p,t)=&~  \langle e^{-iky(t)-ipW(t)}\phi_0(kg(y(t),t)) \dot{S}(t) \rangle \\
    &-ik \langle e^{-iky(t)-ipW(t)}f(y(t),t) \dot{S}(t) \rangle \\
    &-ip \langle e^{-iky(t)-ipW(t)}U(y(t)) \rangle  \\
    =&:~ Q_1+Q_2+Q_3.
\end{split}
\end{equation}

It is obvious that the inverse Fourier transform ($k\rightarrow y$) of $Q_3$ is $-ipU(y)G(y,p,t)$. But for $Q_1$ and $Q_2$, they look a little bit difficult due to the new term $\dot{S}(t)$ compared with \eqref{dGkpt2}. Note that the angle bracket in $Q_1$ denotes the average of two kinds of independent stochastic processes with the joint PDF $G(y(t),W(t),t)$ and L\'{e}vy $\alpha$-stable noise $\theta(t)$ on which $S(t)$ depends.
To deal with the term $Q_1$, we first add a technical delta function $\delta(y-y(t))$ in it and get
\begin{equation*}
  Q_1= \int_{-\infty}^\infty e^{-iky}\phi_0(kg(y,t)) \langle e^{-ipW(t)} \delta(y-y(t)) \dot{S}(t) \rangle  dy.
\end{equation*}
Then applying the technique in \cite{CairoliBaule:2017} of rewriting the functional $W(t)$ as a subordinated process:
\begin{equation*}
  W(t)=V(S(t)),   \qquad  V(s)=\int_0^s U(x(s'))\theta(s')ds'.
\end{equation*}
Substituting $y(t)=x(S(t))$ and $W(t)=V(S(t))$ into $Q_1$ gives the middle term of $Q_1$ as
\begin{equation}\label{Q1B}
  \begin{split}
    & \langle e^{-ipV(S(t))} \delta(y-x(S(t))) \dot{S}(t) \rangle  \\
    &=\int_0^\infty  \langle e^{-ipV(s)} \delta(y-x(s))\delta(t-T(s)) \rangle ds.
  \end{split}
\end{equation}
Taking Laplace transform ($t\rightarrow u$) of \eqref{Q1B}, we obtain
  \begin{equation}\label{Q1uB}
  \begin{split}
      Q_1(u) =&~ \int_{-\infty}^\infty e^{-iky}\phi_0(kg(y,t))   \\
      &\cdot\int_0^\infty \langle e^{-ipV(s)-uT(s)} \delta(y-x(s)) \rangle ds dy.
  \end{split}
  \end{equation}
On the other hand, $G(y,p,t)$ can be rewritten as:
\begin{equation*}
  \begin{split}
    G(y,p,t)&= \langle e^{-ipV(S(t))}\delta(y-x(S(t)))  \rangle    \\
    &= \int_0^\infty  \langle e^{-ipV(s)}\delta(s-S(t))\delta(y-x(s)) \rangle  ds.
  \end{split}
\end{equation*}
So its Laplace transform ($t\rightarrow u$) is
\begin{equation}\label{Gypu}
G(y,p,u)=  \int_0^\infty  \langle e^{-ipV(s)-uT(s)} \theta(s)\delta(y-x(s)) \rangle  ds.
\end{equation}
The characteristic function of the L\'{e}vy process $T(s)$ in \eqref{LEmodelB} is
\begin{equation*}
    \langle e^{-uT(s)}\rangle=e^{-su^\alpha},
\end{equation*}
which yields an important equality in \cite{CairoliBaule:2017} from \eqref{Gypu}:
\begin{equation}\label{GkpuB}
\begin{split}
    G(y,p,u)=&[u+ipU(y)]^{\alpha-1}  \\
    &\cdot \int_0^\infty  \langle e^{-ipV(s)-uT(s)} \delta(y-x(s)) \rangle  ds.
\end{split}
\end{equation}
Comparing \eqref{Q1uB} with \eqref{GkpuB}, we find that
\begin{equation*}
  Q_1(u)= \int_{-\infty}^\infty e^{-iky}\phi_0(kg(y,t)) [u+ipU(y)]^{1-\alpha}G(y,p,u)  dy.
\end{equation*}
Taking the inverse Laplace transform ($u\rightarrow t$), we obtain
\begin{equation}\label{Q_1B}
  Q_1= \int_{-\infty}^\infty e^{-iky}\phi_0(kg(y,t)) \mathcal{D}_t^{1-\alpha} G(y,p,t)  dy.
\end{equation}
As for $Q_2$, it can be obtained by the procedure similar to $Q_1$, i.e.,
\begin{equation}\label{Q_2B}
  Q_2= -ik\int_{-\infty}^\infty e^{-iky} f(y,t) \mathcal{D}_t^{1-\alpha} G(y,p,t)  dy.
\end{equation}
Finally, substituting \eqref{Q_1B} and \eqref{Q_2B} into \eqref{dGkpt2B}, we obtain the forward Feynman-Kac equation in Fourier space:
\begin{equation*}
\begin{split}
    &\frac{\partial G(k,p,t)}{\partial t}= \mathcal{F}_y\{\phi_0(kg(y,t)) \mathcal{D}_t^{1-\alpha} G(y,p,t) \}  \\
    &~~-\mathcal{F}_y\left\{\frac{\partial}{\partial y}f(y,t) \mathcal{D}_t^{1-\alpha} G(y,p,t) + ipU(y)G(y,p,t) \right\}.
\end{split}
\end{equation*}

\section{Backward Feynman-Kac equation with multiplicative noise}\label{AppA}

If $g(x)$ is not a constant in (\ref{LEmodelBe}), then the Fourier transform of $\langle T_\eta\rangle$ becomes
\begin{equation*}
  \mathcal{F}_y\{\langle T_\eta \rangle\}=\left\langle \int_{-\infty}^{\infty}e^{-ik_0x(\tau)}T_\eta e^{ik_0(f(x_0)\tau+g(x_0)\eta(\tau))}dx_0  \right\rangle.
\end{equation*}
We turn $dx_0$ into $dx(\tau)$ and get
\begin{widetext}
\begin{equation}\label{FTeta2}
  \begin{split}
    \mathcal{F}_{x_0}\{\langle T_\eta \rangle\}
    =&~\Big\langle \int_{-\infty}^{\infty}e^{-ik_0x(\tau)}T_\eta e^{ik_0(f(x_0)\tau+g(x_0)\eta(\tau))}dx(\tau)  \Big\rangle
    -\Big\langle \int_{-\infty}^{\infty}e^{-ik_0x(\tau)}T_\eta e^{ik_0(f(x_0)\tau+g(x_0)\eta(\tau))}\frac{d f(x_0)}{d x_0}\tau dx_0  \Big\rangle  \\
    &-\Big\langle \int_{-\infty}^{\infty}e^{-ik_0x(\tau)}T_\eta e^{ik_0(f(x_0)\tau+g(x_0)\eta(\tau))}\frac{d g(x_0)}{d x_0}\eta(\tau) dx_0  \Big\rangle.
  \end{split}
\end{equation}
\end{widetext}
Letting $\tau\rightarrow0$, the second term of $\mathcal{F}_{x_0}\{\langle T_\eta \rangle\}$ is the same as (\ref{ReverseOperator}), i.e.,
\begin{equation}\label{SecondTerm}
  -\tau\mathcal{F}_{x_0}\Big\{\frac{\partial f(x_0)}{\partial x_0}G_{x_0}(p,t)\Big\}.
\end{equation}
Though $\eta(\tau)\rightarrow0$ as $\tau\rightarrow0$, how fast it tends to $0$ is not specific, which arises the challenge of dealing with the first and third terms (\ref{FTeta2}).
To make this point clear, we should define
\begin{equation*}
  M_n(k,\tau)=\langle e^{-ik\eta(\tau)}\eta^n(\tau)\rangle.
\end{equation*}
When $n=0$, $M_0$ is the characteristic function of $\eta(\tau)$, given in \eqref{NoiseChar}. When $n\geq1$, $M_n\rightarrow0$ as $\tau\rightarrow0$. For the case of Gaussian white noise: $M_0=e^{-\tau k^2}$, by some calculations, we have, as $\tau\rightarrow0$,
\begin{equation}\label{M12}
\begin{split}
    &M_1\sim -2ik\tau, \qquad   M_2\sim  2\tau, \\
    &M_n\sim  \tau^2 \sim0 \qquad \forall~n\geq3;
\end{split}
\end{equation}
since $M_n\,(n\geq3)$ are all higher order terms of $\tau$, it can be omitted. But for L\'{e}vy $\beta$-stable noise,  $M_0=e^{-\tau |k|^\beta}$ and all $M_n\,(n\geq1)$ contain the first order term of $\tau$. Here we focus on the case that $\eta(\tau)$ is Gaussian white noise, and use the property \eqref{M12} to deal with the first and third terms in \eqref{FTeta2}.

\begin{widetext}
Denoting the first term as $T_1$ for convenient and using $e^{ik_0f(x_0)\tau}\simeq 1+ik_0f(x_0)\tau$ as before, we get
\begin{equation}\label{FirstTerm1}
       T_1= \Big\langle \int_{-\infty}^{\infty}e^{-ik_0x(\tau)}T_\eta e^{ik_0g(x_0)\eta(\tau)}dx(\tau)  \Big\rangle
        +ik_0\tau \Big\langle \int_{-\infty}^{\infty}e^{-ik_0x(\tau)}T_\eta f(x(\tau))dx(\tau)  \Big\rangle,
\end{equation}
where the latter term equals to
\begin{equation*}
  \tau\mathcal{F}_{x_0}\Big\{\frac{\partial}{\partial x_0}f(x_0)G_{x_0}(p,t)\Big\}.
\end{equation*}
Turn $g(x_0)$ into $g(x(\tau))$ in \eqref{FirstTerm1} by Taylor expansion $g(x_0)=g(x(\tau))+R_g$, where
\begin{equation*}
\begin{split}
    R_g=-(f(x_0)\tau+g(x_0)\eta(\tau))g'(x(\tau))
    +\frac{1}{2}(f(x_0)\tau+g(x_0)\eta(\tau))^2g''(x(\tau))+\cdots,
\end{split}
\end{equation*}
and here for convenient, we use the notation $'$ to denote the first order derivative.
Then the former term of \eqref{FirstTerm1}, denoted as $T_{11}$, becomes
\begin{equation}\label{FirstTerm2}
  T_{11}=\Big\langle \int_{-\infty}^{\infty}e^{-ik_0x(\tau)}T_\eta e^{ik_0g(x(\tau))\eta(\tau)} e^{ik_0R_g\eta(\tau)}  dx(\tau)  \Big\rangle,
\end{equation}
where $e^{ik_0R_g\eta(\tau)}=1+ik_0R_g\eta(\tau)+\cdots$.   Considering $M_n\sim0\,(n\geq3)$ in \eqref{M12}, the second term of $R_g$ and all the latter terms can be omitted since these terms contribute to $\eta^2(\tau)$ and then yields $M_n\,(n\geq3)$ when substituted into \eqref{FirstTerm2}. Therefore, we have
\begin{equation*}
\begin{split}
    T_{11}&=\Big\langle \int_{-\infty}^{\infty}e^{-ik_0x(\tau)}T_\eta e^{ik_0g(x(\tau))\eta(\tau)} e^{-ik_0g(x_0)g'(x(\tau))\eta^2(\tau)}  dx(\tau)  \Big\rangle  \\
    &=\Big\langle \int_{-\infty}^{\infty}e^{-ik_0x(\tau)}T_\eta e^{ik_0g(x(\tau))\eta(\tau)}   dx(\tau)  \Big\rangle
    -ik_0 \Big\langle \int_{-\infty}^{\infty}e^{-ik_0x(\tau)}T_\eta e^{ik_0g(x(\tau))\eta(\tau)} g(x_0)g'(x(\tau))\eta^2(\tau)  dx(\tau)  \Big\rangle  \\
    &=\mathcal{F}_{x_0}\{\langle e^{ik_0g(x_0)\eta(\tau)}\rangle G_{x_0}(p,t)\} -ik_0\mathcal{F}_{x_0}\{\langle e^{ik_0g(x_0)\eta(\tau)}\eta^2(\tau)\rangle g(x_0)g'(x_0)G_{x_0}(p,t)\},
\end{split}
\end{equation*}
where we replace $g(x_0)$ by $g(x(\tau))$ in the latter term and omit the high order term $M_3$. Substituting $M_2$ in \eqref{M12} and $T_{11}$ into \eqref{FirstTerm1} gives
\begin{equation}\label{FirstTerm}
  T_1=\mathcal{F}_{x_0}\{\langle e^{ik_0g(x_0)\eta(\tau)}\rangle G_{x_0}(p,t)\}-2ik_0\tau\mathcal{F}_{x_0}\{g(x_0)g'(x_0)G_{x_0}(p,t)\}
  +\tau\mathcal{F}_{x_0}\Big\{\frac{\partial}{\partial x_0}f(x_0)G_{x_0}(p,t)\Big\}.
\end{equation}
Similarly, still using the property $M_n\sim0,\,n\geq3$, we get the third term in \eqref{FTeta2},
\begin{equation}\label{ThirdTerm}
  T_3= -2\tau \mathcal{F}_{x_0}\Big\{g(x_0)g'(x_0)\frac{\partial G_{x_0}(p,t)}{\partial x_0}\Big\}.
\end{equation}
Combining \eqref{FirstTerm}, \eqref{ThirdTerm} and \eqref{SecondTerm}, we finally get
\begin{equation}\label{FTeta}
\begin{split}
    \mathcal{F}_{x_0}\{\langle T_\eta \rangle\}=&~\mathcal{F}_{x_0}\{\langle e^{ik_0g(x_0)\eta(\tau)}\rangle G_{x_0}(p,t)\}
      -2\tau\mathcal{F}_{x_0}\Big\{\frac{\partial}{\partial x_0}g(x_0)g'(x_0)G_{x_0}(p,t)\Big\}    \\
      &+\tau\mathcal{F}_{x_0}\Big\{f(x_0)\frac{\partial G_{x_0}(p,t)}{\partial x_0}\Big\}
      -2\tau \mathcal{F}_{x_0}\Big\{g(x_0)g'(x_0)\frac{\partial G_{x_0}(p,t)}{\partial x_0}\Big\}.
\end{split}
\end{equation}
Using the characteristic function of L\'{e}vy noise \eqref{NoiseChar} leads to
\begin{equation}\label{FTeta3}
  \langle e^{ik_0g(x_0)\eta(\tau)}\rangle-1 \simeq \tau \phi_0(-k_0g(x_0))=-\tau k_0^2g^2(x_0)  \qquad  \textrm{as} ~\tau\rightarrow0.
\end{equation}
Combining \eqref{FTeta} and \eqref{FTeta3}, by some calculations, we obtain
\begin{equation*}
  \mathcal{F}_{x_0}\{\langle T_\eta \rangle\}-\mathcal{F}_{x_0}\{G_{x_0}(p,t)\}= \tau \mathcal{F}_{x_0}\Big\{g^2(x_0)\frac{\partial^2G_{x_0}(p,t)}{\partial x_0^2}\Big\}
            +\tau\mathcal{F}_{x_0}\Big\{f(x_0)\frac{\partial G_{x_0}(p,t)}{\partial x_0}\Big\}.
\end{equation*}
Substituting this formula into \eqref{BFKE_k1}, dividing \eqref{BFKE_k1} by $\tau$, taking the limit $\tau\rightarrow0$ and making the inverse Fourier transform ($k_0\rightarrow x_0$),
we obtain the backward Feynman-Kac equation:
\begin{equation}
  \frac{\partial G_{x_0}(p,t)}{\partial t}= g^2(x_0)\frac{\partial^2G_{x_0}(p,t)}{\partial x_0^2}
            + f(x_0)\frac{\partial G_{x_0}(p,t)}{\partial x_0}  - ipU(x_0)G_{x_0}(p,t).
\end{equation}
\end{widetext}

%



\bibliographystyle{apsrev4-1}
\bibliography{Reference}

\end{document}